\documentclass[twocolumn,aps,showpacs,superscriptaddress,longbibliography]{revtex4-1}
\usepackage{graphicx}
\usepackage{latexsym}

\begin{document}

\title{Compound twin beams without the need of genuine photon-number-resolving
detection}

\author{Jan Pe\v{r}ina Jr.}
\email{jan.perina.jr@upol.cz} \affiliation{Joint Laboratory of Optics of
Palack\'{y} University and Institute of Physics of the Czech Academy of
Sciences, Faculty of Science, Palack\'{y} University, 17. listopadu 12, 77146
Olomouc, Czech Republic}

\author{Anton\'\i n \v{C}ernoch}
\affiliation{Joint Laboratory of Optics of Palack\'{y} University and Institute
of Physics of the Czech Academy of Sciences, Faculty of Science, Palack\'{y}
University, 17. listopadu 12, 77146 Olomouc, Czech Republic}

\author{Jan Soubusta}
\affiliation{Institute of Physics of the Czech Academy of Sciences, Joint
Laboratory of Optics of Palack\'{y} University and Institute of Physics of CAS,
17. listopadu 50a, 772 07 Olomouc, Czech Republic}

\begin{abstract}
The scheme for building stronger multi-mode twin beams from a greater number of
identical twin beams sufficiently weak so that single-photon sensitive on/off
detectors suffice in their detection is studied. Statistical properties of
these compound twin beams involving the non-classicality are analyzed for
intensities up to hundreds of photon pairs. Their properties are compared with
those of the genuine twin beams that require photon-number-resolving detectors
in their experimental investigations. The use of such compound twin beams for
the generation of sub-Poissonian light and measurement of absorption with
sub-shot-noise precision is analyzed. A suitable theoretical model for the
compound twin beams is developed to interpret the experimental data.
\end{abstract}

\maketitle

\section{Introduction}

In the history of detection of optical fields, two milestones occur. The first
milestone was reached when the solid-state light detectors became sensitive
enough to recognize individual detected photons (due to the used avalanche
photodiodes (APDs) \cite{McIntyre1961}). The second milestone is connected with
the ability to count photons which is the basic method for characterizing
optical fields \cite{Mandel1995,Perina1991}. The first photon-number-resolving
detectors (PNRDs) were simple: Two single-photon counting modules were used to
monitor the outputs of a balanced beam splitter and resolve up to two photons.
More complex geometries were applied \cite{Paul1996} including the so-called
fiber-loop PNRDs \cite{Achilles2003,Fitch2003,Haderka2004} with time
multiplexing to extend the range of detected photons. Spatial multiplexing
exploited in intensified CCD (iCCD) cameras \cite{Haderka2005a,PerinaJr2012}
and silicon matrix photodetectors \cite{Chesi2019} considerably enlarge the
detectable photon numbers. Finally, genuine PNRDs based on super-conducting
bolometers \cite{Harder2016,MaganaLoaiza2019} were constructed. In such PNRDs,
the number of detected photons is linearly proportional to the energy absorbed
and indirectly detected in super-conducting metallic wires \cite{Miller2003}.

The use of PNRDs in the investigation of nonclassical properties of optical
fields including the entangled ones revealed various states with unusual
physical properties: states with the reduced photon-number fluctuations
(sub-Poissonian states \cite{PerinaJr2017c}, the Fock states as extreme
examples), states containing only odd or even photon numbers
\cite{Malkin1979,Gerry1993}, states with perfect correlations in photon numbers
[twin beams (TWBs)] \cite{Friberg1985}, to name few. These states have found
their application in metrology (absolute detector calibration
\cite{Klyshko1980,Migdall1999,Brida2010,PerinaJr2014a}, quantum imaging
\cite{Brida2010a,Genovese2016} as well as sub-shot-noise measurements
\cite{Losero2018}). They have also been considered in quantum communications
\cite{Saleh1987}. However, the PNRDs at present are complex and expensive
devices and this limits the use of the above methods and applications.

This brings us to the question whether the PNRDs with their capabilities to
detect optical fields can somehow be substituted by simpler APDs with their
on/off detection, similarly as it is done in time and spatially multiplexed
PNRDs. Such substitution is in principle possible for multi-mode optical fields
that are composed of a greater number of spatial and spectral modes. When the
mean numbers of photons in individual modes are considerably lower than one we
may monitor the fields in individual modes by APDs. Provided that we are able
to prepare the fields of individual modes independently, we can detect their
properties by APDs mode by mode. Similarly as the whole optical field is built
from individual modes its photon-number distribution is composed of the
contributions from APDs that monitor individual modes. Whereas such approach is
meaningless for the analysis of unknown optical fields, it is prospective for
applications that use optical fields with well-defined properties, as mentioned
above.

\begin{figure}  
 \includegraphics[width=0.95\hsize]{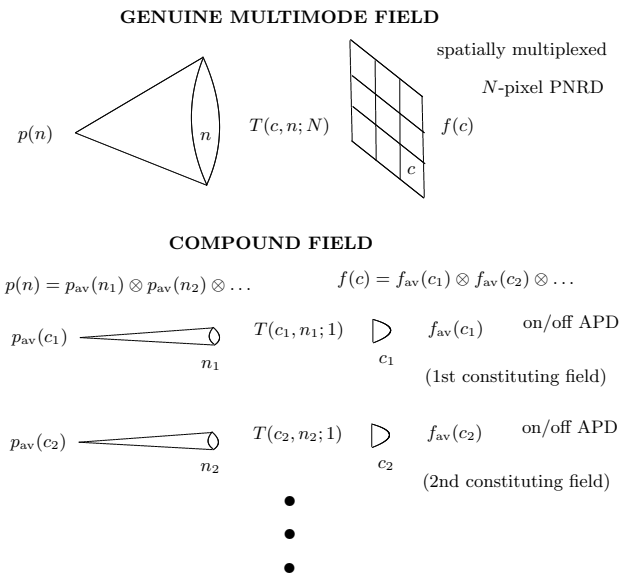}
 \caption{Substitution scheme for a stronger multi-mode field with a photon-number distribution $ p(n) $ detected
  by an $ N $-pixel spatially multiplexed PNRD giving a photocount histogram $ f(c) $.
  The stronger field is replaced by $ N $ identical weak fields with
  photon-number distributions $ p_{\rm av}(n) $ detected by APDs that provide
  photocount histograms $ f_{\rm av}(c) $. The photon-number distribution $
  p(n) $ and photocount histogram $ f(c) $ are obtained by $ N $-fold
  convolution from the photon-number distributions $ p_{\rm av}(n) $ and
  photocount histograms $ f_{\rm av}(c) $. The elements of detection matrix $ T(c,n;N) $
  give the probabilities of detecting $ c $ photocounts out of $ n $ impinging
  photons by an $ N $-pixel detector.}
\label{fig1}
\end{figure}

To demonstrate this approach, we consider TWBs and their application in quantum
imaging. Several mean photon pairs per an imaged pixel are typically needed.
The used TWB is typically composed of say $ N $ independent spatial modes and
its statistical operator $ \hat{\varrho} $ is written as a product of
statistical operators $ \hat{\varrho}_j $ of individual modes:
\begin{equation}  
 \hat{\varrho} = \prod_{j=1}^N \hat{\varrho}_j.
\label{1}
\end{equation}
In the considered TWB, individual modes exhibit perfect photon pairing
described by the thermal distribution with $ b_{{\rm p,j}} $ mean photon pairs,
\begin{equation}  
 \hat{\varrho}_j = \sum_{n_{\rm p}=0}^\infty \frac{ b_{{\rm p},j}^{n_{\rm p}} }{
  (1 + b_{{\rm p},j})^{ n_{\rm p}+1} } |n_{\rm p}\rangle_{{\rm s}_j}
  |n_{\rm p}\rangle_{{\rm i}_j} {}_{{\rm i}_j}\langle n_{\rm p}|
  {}_{{\rm i}_j}\langle n_{\rm p}|,
\label{2}
\end{equation}
and the Fock state $ |n_{\rm p}\rangle_{{\rm s}_j} $ ($ |n_{\rm p}\rangle_{{\rm
i}_j} $) contains $ n_{\rm p} $ photons in the signal (idler) field of mode $ j
$. The mean photon-pair numbers $ b_{{\rm p},j} $ of modes differ, but this has
practically no influence on the properties exploited in the TWB applications.
For this reason, we may substitute the TWB with statistical operator $
\hat{\varrho} $ in Eq.~(\ref{1}) by a TWB composed of $ N $ equally populated
individual modes. Such TWB is described by the following statistical operator $
\hat{\varrho}^{\rm c} $
\begin{equation}   
 \hat{\varrho}^{\rm c} = \hat{\varrho}_{\rm av}^{\oplus N}
\label{3}
\end{equation}
in which the statistical operator $ \hat{\varrho}_{\rm av} $ given by
Eq.~(\ref{2}) contains $ b_{\rm p} = \sum_{j=1}^N b_{{\rm p},j} / N $ mean
photon pairs. The TWB with statistical operator $ \hat{\varrho}^{\rm c} $ in
Eq.~(\ref{3}) is then suitable for practical realization in which we prepare $
N $ times the field with statistical operator $ \hat{\varrho}_{\rm av} $. Thus,
instead of measuring a stronger TWB given in Eq.~(\ref{1}) by two PNRDs, we
measure $ N $ times a weak TWB with statistical operator $ \hat{\varrho}_{\rm
av} $ by two APDs and sum the results to arrive at those appropriate to a TWB
given in Eq.~(\ref{3}). We call such TWBs the compound TWBs. The detection
scheme is illustrated in Fig.~\ref{fig1} considering one of the beams that
compose a TWB. We note that similar scheme was applied in the experimental
study of non-Gaussian states in Ref.~\cite{Straka2018}.

The compound TWBs generated in the substitution scheme are interesting on their
own. Here, we demonstrate their usefulness when studying the behavior of TWB
non-classicality for the intensities varying by three orders in magnitude. The
substitution scheme assures close similarity of parameters of compound TWBs
that differ in their intensities by orders in magnitude. Such analysis would
not be possible for the usual experimental TWBs.

Sub-shot-noise imaging and quantum metrology represent the most important
applications of stronger TWBs (composed of from several to several hundreds of
photon pairs) at present. In one of their variants, they use sub-Poissonian
fields as light sources for monitoring absorption coefficients of samples. They
rely on the fact that the fluctuations in photon numbers in such fields are
suppressed below the shot-noise-limit which allows to reach better measurement
precision. Post-selection from a TWB
\cite{Rarity1987,Laurat2003,PerinaJr2013b,PerinaJr2017c} whose one arm is
monitored by a PNRD belongs to the most efficient methods of
sub-Poissonian-light generation. Using the above discussed similarity of
parameters of composed TWBs, we investigate the non-classicality of
sub-Poissonian fields originating in the compound TWBs as it depends on the
fields' intensity.

Finally, as the third example of the use of compound TWBs, we directly analyze
the performance of compound TWBs in sub-shot-noise measurements
\cite{Jakeman1986,Giovanetti2006,Brida2010a,Losero2018} by determining the
precision of absorption measurements as it depends on the compound-TWB
intensity. Such measurements were recently applied in microscopy
\cite{Li2018,SabinesChesterkind2019} and spectroscopy \cite{Whittaker2017}.

We note that the use of APDs instead of genuine PNRDs in the discussed
substitution scheme also benefits from considerably higher absolute detection
efficiencies of APDs.

The paper is organized as follows. Photon-number as well as photocount
distributions of genuine and compound TWBs are introduced in Sec.~II. In
Sec.~III, the experiment is described and analyzed and the main characteristics
of the generated fields are given. The determination of effective detection
efficiency, which is an important parameter of PNRDs, is discussed in Sec.~IV.
Non-classicality of compound TWBs is analyzed in Sec.~V using suitable
non-classicality identifiers and the accompanying non-classicality depths. The
generation of sub-Poissonian light by post-selection from compound TWBs is
discussed in Sec.~VI. Sub-shot-noise measurements of absorption coefficients
using the compound TWBs are investigated in Sec.~VII. Conclusions are drawn in
Sec.~VIII. In appendix, the model for describing the correlations in intra
pump-pulse intensity fluctuations is developed.

\section{Photon-number and photocount distributions of genuine and compound twin beams.}

The joint photon-number distribution $ p_{\rm si}(n_{\rm s},n_{\rm i}) $ of an
arbitrary TWB generated from parametric down-conversion can be described as a
two-fold convolution of three photon-number distributions $ p_{\rm p} $, $
p_{\rm s} $ and $ p_{\rm i} $ belonging in turn to the paired, noise signal and
noise idler components of the described TWB \cite{PerinaJr2013a}:
\begin{equation}  
 p_{\rm si}(n_{\rm s},n_{\rm i}) = \sum_{n=0}^{{\rm min}(n_{\rm s},n_{\rm i})}
  p_{\rm s}(n_{\rm s}-n) p_{\rm i}(n_{\rm i}-n) p_{\rm p}(n);
\label{4}
\end{equation}
$ n_{\rm s} $ ($ n_{\rm i} $) gives the number of signal (idler) photons. In
the model, we assume the photon-number distributions of the components
constituting the TWB in the form of the multi-mode thermal Mandel-Rice
distribution \cite{Perina1991} that is derived for a field with $ M_a $ equally
populated modes and mean photon (-pair) number $ B_a $ per mode:
\begin{equation}  
  p_a(n;M_a,B_a) = \frac{\Gamma(n+M_a) }{n!\, \Gamma(M_a)}
  \frac{B_a^n}{(1+B_a)^{n+M_a}}, \hspace{5mm} a={\rm s,i,p}
\label{5}
\end{equation}
and $ \Gamma $ denotes the $ \Gamma $-function.

The joint photocount distribution $ f_{\rm si}(c_{\rm s},c_{\rm i}) $
\cite{Perina1991} registered by two PNRDs composed of $ N_s $ and $ N_i $
equally illuminated pixels with on/off detection by the field with
photon-number distribution $ p_{\rm si}(n_{\rm s},n_{\rm i}) $ given in
Eq.~(\ref{4}) is given as follows:
\begin{equation} 
 f_{\rm si}(c_{\rm s},c_{\rm i}) = \sum_{n_{\rm s},n_{\rm i}=0}^{\infty}
  T_{\rm s}(c_{\rm s},n_{\rm s};N_{\rm s}) T_{\rm i}(c_{\rm i},n_{\rm i};N_{\rm i})
  p_{\rm si}(n_{\rm s},n_{\rm i});
\label{6}
\end{equation}
$ c_{\rm s} $ [$ c_{\rm i} $] denotes the number of signal (idler) photocounts
that denote the photoelectrons registered by a detector after the absorption of
photons.

For a detector with quantum detection efficiency $ \eta_a $, number $ N_a $ of
pixels and mean dark count number per pixel $ D_a $, the elements of detection
matrices $ T_a $, $ a ={\rm s,i} $, introduced in Eq.~(\ref{6}) are derived in
the form~\cite{PerinaJr2012}:
\begin{eqnarray}     
  T_a(c_a,n_a;N_a) &=& \left(\begin{array}{c} N_a \\ c_a \end{array}\right) (1-D_a)^{N_a}
   (1-\eta_a)^{n_a} (-1)^{c_a} \nonumber \\
  & &  \mbox{} \hspace{-18mm} \times  \sum_{l=0}^{c_a} \left(\begin{array}{c} c_a \\ l \end{array}\right)
    \frac{(-1)^l}{(1-D_a)^l}  \left( 1 + \frac{l}{N_a} \frac{\eta_a}{1-\eta_a}
   \right)^{n_a}.
\label{7}
\end{eqnarray}
The derivation of Eq.~(\ref{7}) shows that when the mean number of photons
illuminating one pixel is much less than one, i.e. when we can neglect
illumination of a pixel by more than one photon, the elements of detection
matrix $ T_a(c_a,n_a;N_a) $ in Eq.~(\ref{7}) can be written as composed of the
elements of independent detection matrices $ T_a(c_a,n_a;1) $ characterizing
the detection by individual pixels.

In the substitution scheme, we assume that a genuine stronger TWB is replaced
by a compound TWB, i.e. an ensemble of $ N $ identical constituting TWBs
detected by on/off APDs such that the mean photon numbers of the genuine and
compound TWBs are the same. Though the mean photon-pair numbers of the
constituting TWBs are much less than one, we still have to consider their
multi-mode structure (see the discussion below). Provided that we have $ m_a $
modes in a constituting TWB, the mean photon (-pair) numbers $ b_a $ per mode
of a constituting TWB are given as $ b_a = B_aM_a/(m_aN) $, $ a={\rm s,i,p} $.
The formula analogous to that in Eq.~(\ref{4}) gives us the corresponding joint
signal-idler photon-number distribution $ p^w_{\rm si}(n_{\rm s},n_{\rm i}) $
of a constituting TWB. The constituting TWBs are detected by two APDs whose
operation is described by the elements of detection matrix $ T_a $ in
Eq.~(\ref{7}) assuming $ N_a = 1 $, $ a = {\rm s,i} $. The joint photocount
distribution $ f^w_{\rm si}(c_{\rm s},c_{\rm i}) $ appropriate to one
constituting TWB is given as:
\begin{equation} 
 f^w_{\rm si}(c_{\rm s},c_{\rm i}) = \sum_{n_s,n_i=0}^{\infty} T_s(c_s,n_s;1) T_i(c_i,n_i;1)
  p^w_{\rm si}(n_{\rm s},n_{\rm i}).
\label{8}
\end{equation}

For a compound TWB formed by $ N $ constituting TWBs, the probability $ f^{\rm
c}_{\rm si} $ of having $ c_{\rm s} $ signal photocounts together with $ c_{\rm
i} $ idler photocounts is determined by the following multiple convolution
\begin{eqnarray}   
 f^{\rm c}_{\rm si}(c_{\rm s},c_{\rm i};N) &=& \sum_{c_{{\rm s},1},c_{{\rm i},1}=0}^{1}
  \cdots \sum_{c_{{\rm s},N},c_{{\rm i},N}=0}^{1} \delta_{c_{\rm s}, \sum_{j=1}^{N} c_{{\rm s},j}}
  \nonumber \\
 & & \delta_{c_{\rm i}, \sum_{j=1}^{N} c_{{\rm i},j}}
   \prod_{j=1}^{N} f^w_{\rm si}(c_{{\rm s},j},c_{{\rm i},j})
\label{9}
\end{eqnarray}
in which $ \delta $ stands for the Kronecker symbol.

In analogy to Eqs.~(\ref{8}) and (\ref{9}), the probability $ p_{c,{\rm
i}}^w(n_{\rm i};c_{\rm s}) $ of having $ n_{\rm i} $ idler photons conditioned
by the detection of $ c_{\rm s} $ signal photocounts in a constituting TWB is
given as:
\begin{equation} 
 p^w_{c,{\rm i}}(n_{\rm i};c_{\rm s}) = \sum_{n_{\rm s}=0}^{\infty} T_{\rm s}(c_{\rm s},n_{\rm s};1)
  p^w_{\rm si}(n_{\rm s},n_{\rm i}).
\label{10}
\end{equation}
The conditional probability $ p_{c,{\rm i}}^{\rm c}(n_{\rm i};c_{\rm s}) $ of
having $ n_{\rm i} $ idler photons after detecting $ c_{\rm s} $ signal
photocounts in a compound TWB formed by $ N $ constituting TWBs is then
expressed as:
\begin{eqnarray}   
 p_{c,{\rm i}}^{\rm c}(n_{\rm i};c_{\rm s},N) &=& \sum_{c_{{\rm s},1}=0}^{1}
  \cdots \sum_{c_{{\rm s},N}=0}^{1} \delta_{c_{\rm s}, \sum_{j=1}^{N} c_{{\rm s},j}}
   \sum_{n_{{\rm i},1}=0}^{n_{\rm i}}\cdots \nonumber \\
 & & \hspace{-2mm} \sum_{n_{{\rm i},N}=0}^{n_{\rm i}}
   \delta_{n_{\rm i}, \sum_{j=1}^{N} n_{{\rm i},j} }
   \prod_{j=1}^{N} p^w_{c,{\rm i}}(n_{{\rm i},j};c_{{\rm s},j}).
\label{11}
\end{eqnarray}

\section{Experimental characterization of compound twin beams}

\begin{figure} 
 \centering
 \includegraphics[width=0.99\hsize]{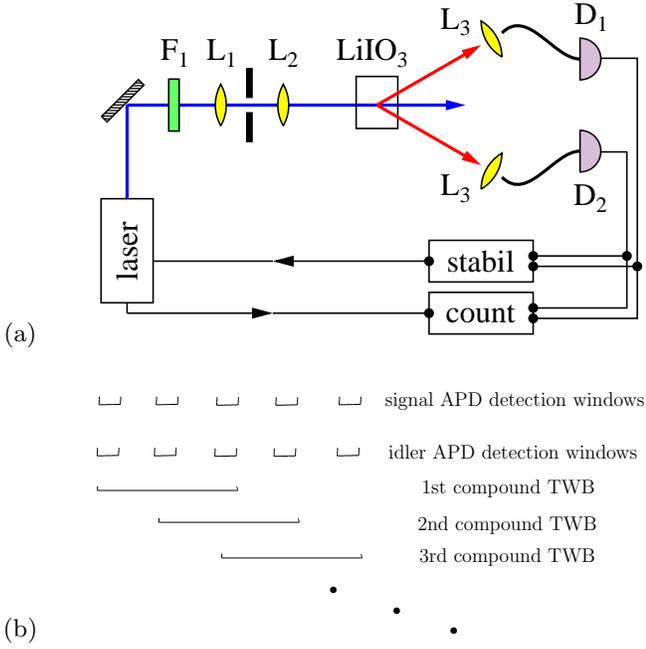}

 \caption{(a) Experimental setup for generation of photon pairs in a nonlinear crystal
  LiIO$ _3 $ by short optical pulses and their detection by single-photon APDs
  D$ _1 $ and D$ _2 $; frequency filter F$_1 $ and lenses L$ _{1,2,3} $ transform the used beams.
  More details are written in the text.
  (b) Signal APD and idler APD detection windows and their grouping into triplets that give the
  measurement on compound TWBs formed by three constituting TWBs.}
\label{fig2}
\end{figure}

The properties of compound TWBs were experimentally investigated in the setup
shown in Fig.~\ref{fig2}.
The process of type-I parametric down-conversion in a LiIO$_3$ nonlinear
crystal pumped by the third harmonic of an Nd-YAG laser at 355~nm and
repetition rate 2.5~kHz was used to generate photon pairs with degenerate
photon wavelengths centered at 710~nm. The laser was kept close above the
threshold to give the power of only a few of mW. The third harmonic was
spectrally cleaned by dichroic mirrors and 10-nm-wide spectral filter. To have
the pump beam with an acceptable spatial profile in the nonlinear crystal we
performed spatial filtering by a $4f$ system composed of two lenses (L$_1$,
L$_2$) with the focal lengths of 20~mm and 50~mm and the 50-$ \mu $m-wide round
aperture. The signal and idler photons were detected by single-photon APDs
(Count-NIR from Laser Components) with nominal detection efficiencies about
80~\% and very low dark count rates ($<50$~Hz). Before being detected, the
photons were spatially filtered by coupling into multi-mode fibers by means of
lenses L$_3$ (15.3~mm focal length, 5~mm clear aperture). Electronic signals
from the detectors were recorded simultaneously with the trigger from the laser
using the counting logic electronics and directly sent to a computer for
recording. Duration of pump pulses was 6.5~ns and similarly the signal and
idler photons were emitted simultaneously in 6.5-ns-long time windows. Time
delay between subsequent pulses was 4~ms. The detection events were recorded
continuously and they formed two synchronized sequences of on/off detections
(in the signal and idler beams). The laser power was stabilized in a feed-back
loop to keep stable operation of the experiment over tens of hours. Feedback
was provided by monitoring the average single-photon detection rate during the
measurement: Whenever the detection rate changed by more than 5~\% the laser
pump power was adjusted to compensate for this declination.

The analyzed experimental data represent a sequence of $ 695 \times 10^6 $
measurements by single-photon APDs in individual detection windows (triggered
by individual pump pulses) with the following four possible outcomes: no
detection at both the signal and idler detectors, detection only at the signal
detector, detection only at the idler detector or the coincidence detection at
both detectors. An outcome in each detection window represents the measurement
on one constituting TWB. Grouping of the outcomes in neighbor detection windows
then gives us the information about compound TWBs. We have grouped the
detections in $ N $ subsequent detection windows to arrive at the experimental
photocount histogram $ f^{\rm c}_{\rm si}(c_{\rm s},c_{\rm i};N) $
characterizing a compound TWB formed by $ N $ constituting TWBs. When forming
the photocount histograms of compound TWBs, the outcomes in individual
detection windows were multiply used following the scheme presented in
Fig.~\ref{fig2}(b).

The joint signal-idler photon-number distribution $ p^{\rm c}_{\rm si}(n_{\rm
s},n_{\rm i};N) $ of a compound TWB formed by $ N $ constituting TWBs has then
been revealed using the reconstruction method based on the maximum-likelihood
(ML) approach \cite{Dempster1977,Vardi1993}. It gives us the following
iteration procedure for the reconstructed photon-number distribution $ p^{\rm
c}_{\rm si} $ ($ j $ stands for the iteration index):
\begin{eqnarray}   
 p^{{\rm c}(j+1)}_{\rm si}(n_{\rm s},n_{\rm i};N)&=& \sum_{c_{\rm s},c_{\rm i}=0}^{\infty}
  F^{(j)}_{\rm si}(c_{\rm s},c_{\rm i};N) T_{\rm s}(c_{\rm s},n_{\rm s};N) \nonumber \\
 & & \hspace{-3mm} \times  T_{\rm i}(c_{\rm i},n_{\rm i};N),
\label{12} \\
 F^{(j)}_{\rm si}(c_{\rm s},c_{\rm i};N) &=& f^{\rm c}_{\rm si}(c_{\rm s},c_{\rm i};N) \Biggl[
  \sum_{n'_{\rm s},n'_{\rm i}=0}^{\infty}  T_{\rm s}(c_{\rm s},n'_{\rm s};N) \nonumber \\
 & & \hspace{-20mm} \times T_{\rm i}(c_{\rm i},n'_{\rm i};N) p^{{\rm c}(j)}_{\rm si}(n'_{\rm s},n'_{\rm i};N) \Bigr]^{-1},
  \hspace{2mm} j=0,1,\ldots \; . \nonumber
\end{eqnarray}

We characterize the obtained compound TWBs by their photocount (photon-number)
moments $ \langle c_{\rm s}^k c_{\rm i}^l \rangle $ ($ \langle n_{\rm s}^k
n_{\rm i}^l \rangle $) determined along the formula:
\begin{equation} 
 \langle c_{\rm s}^k c_{\rm i}^l \rangle = \sum_{c_{\rm s},c_{\rm i}=0}^{N}
  c_{\rm s}^k c_{\rm i}^l f^{\rm c}_{\rm si}(c_{\rm s},c_{\rm i};N).
\label{13}
\end{equation}

Fano factors $ F_{c,a} $ ($ F_{n,a} $) together with mean photocount (photon)
numbers $ \langle c_a \rangle $ ($ \langle n_a \rangle $), $ a = {\rm s, i} $,
defined in Eq.~(\ref{13}) are used to characterize the signal and idler
marginal distributions:
\begin{equation}   
 F_{c,a} = \frac{ \langle (\Delta c_a)^2\rangle }{
   \langle c_a\rangle } ,
\label{14}
\end{equation}
$ \Delta c_a \equiv c_a - \langle c_a\rangle $.

On the other hand, the noise-reduction-parameter $ R_c $ ($ R_n $) and
covariance $ C_{c} $ ($ C_{n} $) of the photocount (photon) numbers $ c_{\rm s}
$ and $ c_{\rm i} $ ($ n_{\rm s} $ and $ n_{\rm i} $) are applied to quantify
both quantum and classical correlations between the signal and idler fields:
\begin{eqnarray}  
 R_{c} &=& \frac{ \langle [\Delta (c_{\rm s}- c_{\rm i})]^2\rangle }{
   \langle c_{\rm s}\rangle + \langle c_{\rm i} \rangle },
\label{15} \\
 C_{c} &=& \frac{ \langle c_{\rm s} c_{\rm i}\rangle }{
  \sqrt{ \langle c_{\rm s}^2\rangle \langle c_{\rm i}^2\rangle }}.
\label{16}
\end{eqnarray}

\begin{figure}  
 \includegraphics[width=0.99\hsize]{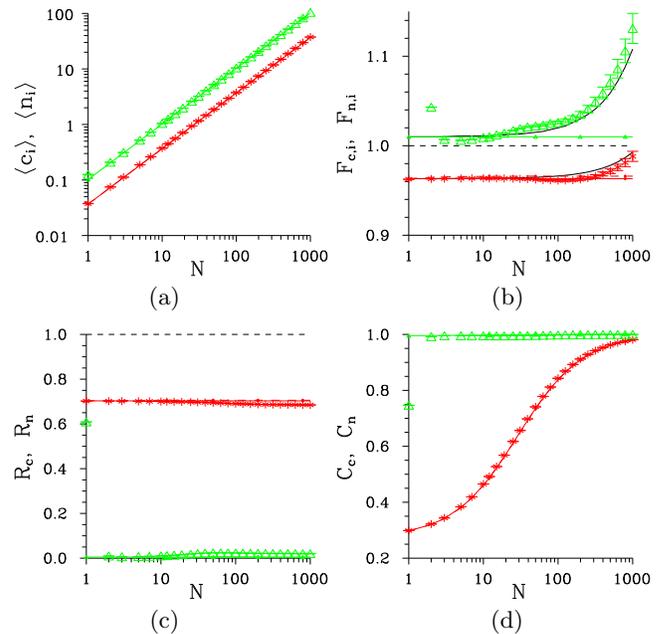}

 \caption{(a) Mean number of photons $ \langle n_{\rm i}\rangle $ (photocounts $ \langle c_{\rm i}\rangle $) and (b)
  Fano factor $ F_{n,{\rm i}} $ ($ F_{c,{\rm i}_1} $) of the idler field, (c) noise-reduction-parameter $ R_{n} $ ($ R_{c} $),
  and (d) covariance $ C_{n} $ ($ C_{c} $) of the compound TWBs as they depend on the number $ N $ of grouped detection windows.
  Isolated symbols with error bars [in (a), (c) and (d) smaller than the plotted symbols] are drawn for the experimental photocount
  histograms (red $ \ast $) and fields reconstructed by the ML approach (green $ \triangle $).
  Solid (dashed) curves with symbols originate in the model of compound (genuine) TWBs detected by APDs ($ N $-pixel PNRDs).
  Black plain solid curves are drawn for the model of compound TWBs with correlations in intra pump-pulse intensity fluctuations.
  In (a), (c) and (d), the red solid and dashed curves nearly coincide.
  The horizontal plain dashed lines indicate the border between the classical and
  non-classical regions ($ F = 1 $, $ R = 1 $).}
\label{fig3}
\end{figure}

The mean photocount numbers $ \langle c_a \rangle $ as well as the
corresponding mean photon numbers $ \langle n_a \rangle $, $ a={\rm s,i} $, of
the marginal signal and idler fields of the compound experimental TWBs
naturally increase linearly with the number $ N $ of constituting TWBs given by
the number of grouped detection windows, as documented in Fig.~\ref{fig3}(a).
The Fano factors $ F_{c,a} $ of the marginal experimental photocount histograms
attain the values slightly smaller than 1 [for the idler Fano factor $
F_{c,{\rm i}} $, see Fig.~\ref{fig3}(b)]. The values of Fano factors $ F_{c,a}
$, $ a={\rm s,i} $, of the detected photocount histograms smaller than 1
originate in the pile-up effect \cite{PerinaJr2012,Chesi2019}. This effect
occurs in the detection with a single-photon APD that allows to register up to
one photocount. As the elements of detection matrix in Eq.~(\ref{7})
incorporates this effect, any reconstruction that uses these elements corrects
for the pile-up effect. This is demonstrated in Fig.~\ref{fig3}(b) where the
values of the idler photon Fano factor $ F_{n,{\rm i}} $ are greater than 1, in
agreement with the classical character of the marginal fields of compound TWBs.
The photon Fano factors $ F_{n} $ of the marginal fields reconstructed by the
ML approach systematically increase with the increasing number $ N $ of grouped
detection windows. This behavior is observed also for the Fano factors $ F_{c}
$ characterizing the photocount histograms, especially for the greater numbers
$ N $ of grouped detection windows [Fig.~\ref{fig3}(b)]. This increase is
caused by weak temporal instability of the pump-pulse intensities that
manifests themselves as correlations in intra pump-pulse intensity
fluctuations. A stochastic model that describes this effect is developed in
Appendix.

We compare the quantities characterizing both the experimental photocount
histograms $ f^{\rm c}_{\rm si}(c_{\rm s},c_{\rm i};N) $ and photon-number
distributions $ p^{\rm c}_{\rm si}(n_{\rm s},n_{\rm i};N) $ reconstructed by
the ML approach with the predictions of two basic models. In the first model of
a compound TWB, we assume that each detection window monitored by two APDs (one
in the signal beam, the other in the idler beam) is illuminated by a Gaussian
constituting TWB with $ m_{\rm p} = m_{\rm s} = m_{\rm i} = 10 $ modes and mean
photon (-pair) numbers per mode $ b_{\rm p} = 1.0185 \times 10^{-2}$, $ b_{\rm
s} = 8 \times 10^{-5}$ and $ b_{\rm i} = 2 \times 10^{-5}$. We made the
assumption $ m_{\rm p} = m_{\rm s} = m_{\rm i} $ for the numbers of modes of
the (typical) constituting TWB and estimated the number $ m_{\rm p} $ of modes
in the dominant paired component of the constituting TWB from the reconstructed
Fano factors $ F_{n_{\rm s}} $ and $ F_{n_{\rm i}} $. Mean photon (-pair)
numbers $ b_{\rm p} $, $ b_{\rm s} $ and $ b_{\rm i} $ were then set to accord
with the mean photon numbers $ \langle n_{\rm s} \rangle $ and $ \langle n_{\rm
i} \rangle $ of the constituting TWB reconstructed by the ML approach. We note
that the constituting TWB is practically noiseless (the unpaired noise photons
comprise less than 1~\% of the constituting TWB intensity). The second model is
simpler. It assumes a genuine strong TWB having $ M_{a} = N m_a $ modes each
containing $ b_a $ mean photons or photon pairs, $ a ={\rm p,s,i} $ and its
monitoring by two genuine PNRDs resolving up to $ N $ photons ($ N $-pixel
PNRDs).

The simpler second model of a genuine TWB detected by two $ N $-pixel PNRDs
with the photon-number distribution $ p_{\rm si}(n_s,n_i) $ given in
Eq.~(\ref{4}) predicts the Fano factors $ F_n $ and $ F_c $ independent of the
number $ N $ of detection pixels (windows) [see the dashed curves in
Fig.~\ref{fig3}(b)]. We arrive at nearly constant values of the photocount Fano
factors $ F_c $ also when we apply the more elaborated first model of the
compound TWB whose constituting TWBs are detected by APDs. The model is
described by Eq.~(\ref{9}) that gives the solid curves in Fig.~\ref{fig3}(b).
The difference between two models is the following. In case of a genuine TWB
and $ N $-pixel PNRDs, each photon, independently on the mode in which it
resides, can be detected by any pixel. It contrasts with the case of a compound
TWB detected by APDs in which photons from a $ k $-th constituting TWB can be
detected only in the $ k $-th detection window of the used APDs ($ k=1,\ldots N
$). Thus, whereas the pile-up effect does not change with the increasing number
$ N $ of detection windows of APDs in case of compound TWBs, it weakens with
the increasing number $ N $ of detection pixels of PNRDs in case of genuine
TWBs. Reduction of the pile-up effect in the latter case then results in the
small increase of the artificially low values of the photocount Fano factors $
F_c $ with the increasing number $ N $ of pixels [see the dashed curve in
Fig.~\ref{fig3}(b)].

The comparison of the results of both models with the measured and
reconstructed values of Fano factors $ F_c $ and $ F_n $ in Fig.~\ref{fig3}(b)
reveals the relatively strong influence of the above mentioned correlations in
intra pump-pulse intensity fluctuations. The corresponding model presented in
Appendix, that represents a suitable generalization of the first model,
provides the solid plain curves in Fig.~\ref{fig3}(b) that are in good
agreement with the measured values. According to the model in Appendix, the
greater the number $ N $ of grouped detection windows is, the greater the
values of Fano factors $ F $ are. However, as the correlations in intra
pump-pulse intensity fluctuations are specific to our experimental setup we
refer to the model in Appendix only when needed.

On the other hand, correlations quantified by the photocount and photon
noise-reduction-parameters $ R_c $ and $ R_n $ depend only weakly on the number
$ N $ of grouped detection windows, as shown in Fig.~\ref{fig3}(c). Whereas the
values of photocount noise-reduction-parameters $ R_c $ are around 0.7 in
agreement with the detection efficiencies of the used APDs, the values of
noise-reduction-parameters $ R_n $ of the compound TWBs reconstructed by the ML
approach are close to 0 indicating nearly ideal pairing of the signal and idler
photons.

Covariances $ C_{c} $ and $ C_{n} $ of photocount and photon numbers,
respectively, are plotted in Fig.~\ref{fig3}(d). Whereas the covariances $
C_{n} $ of the reconstructed compound TWBs are close to the maximal allowed
value 1 (excluding the TWB with $ N=1 $ that is not suitable for ML
reconstruction), the values of photocount covariances $ C_{c} $ gradually
increase with the increasing number $ N $ of grouped detection windows. They
are close to 1 for sufficiently great numbers $ N $ of grouped detection
windows. This behavior follows from a simple model of $ N $ independent
constituting TWBs each being described by the signal and idler photocount
numbers $ c_{{\rm s},j} $ and $ c_{{\rm i},j} $, $ j=1,\ldots,N $. The overall
photocount numbers $ c_a = \sum_{j=1}^N c_{a,j} $, $ a={\rm s,i} $, then have
the moments $ \langle c_{\rm s} c_{\rm i}\rangle = N(N-1) \langle c_{\rm
s}\rangle^w \langle c_{\rm i}\rangle^w + N \langle c_{\rm s}c_{\rm i} \rangle^w
$ and $ \langle c_a^2\rangle = N(N-1) [\langle c_a\rangle^w]^2 + N \langle
c_a^2 \rangle^w $, $ a = {\rm s,i} $, expressed in the moments of individual
constituting TWBs $ \langle c_a\rangle^w $, $ \langle c_a^2\rangle^w $, $
a={\rm s,i} $, and $ \langle c_{\rm s}c_{\rm i} \rangle^w $. Whereas we have $
C_c = [\langle c_{\rm s} c_{\rm i}\rangle^w ]/[ \sqrt{ \langle c_{\rm
s}^2\rangle^w \langle c_{\rm i}^2\rangle^w } ] $ for $ N=1 $ constituting TWB,
$ C_c \rightarrow 1 $ in the limit of large number $ N $ of constituting TWBs.

\begin{figure}  
 \includegraphics[width=0.99\hsize]{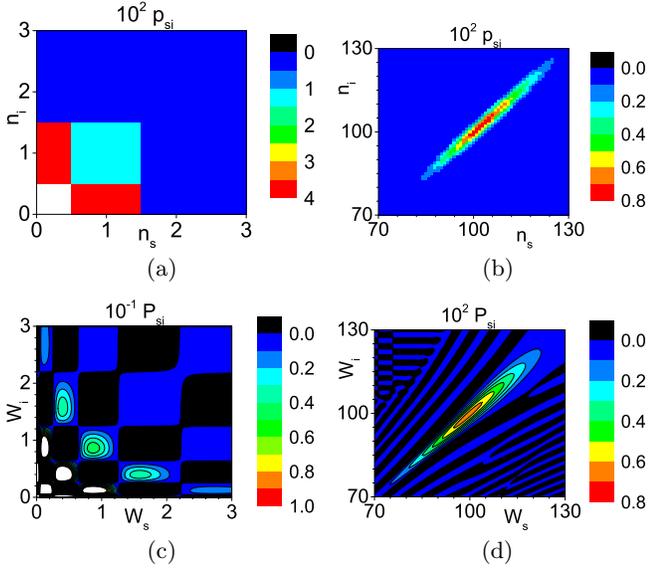}

 \caption{(a,b) Photon-number distribution $ p_{\rm si}(n_{\rm s},n_{\rm i}) $ and (c,d) the corresponding
  quasi-distribution $ P_{{\rm si},s}(W_{\rm s},W_{\rm i}) $ of integrated intensities for $ N=1 $ [(a,c)]
  and $ N=1000 $ [(b,d)].
  In (c) [(d)], $ s= 0.5 $ [$ s= 0 $]. In (a) and (c), the values in white areas are greater than
  those explicitly indicated at the $ z $ scale.}
\label{fig4}
\end{figure}

Intensities of the compound TWBs analyzed in Fig.~\ref{fig3} vary by three
orders in magnitude: The weakest analyzed TWB formed by just one constituting
TWB and detected in one detection window contains 0.102 mean photon pairs, the
strongest compound TWB composed of 1000 constituting TWBs and thus detected in
1000 detection windows is composed of 102 photon pairs. Whereas the
contribution of the vacuum state prevails in the weakest TWB, the joint
photon-number distribution $ p_{\rm si} $ of the strongest analyzed compound
TWB exhibits a well developed pair-wise structure localized around the mean
photon values [compare Figs.~\ref{fig4}(a) and (b)]. These two types of
photon-number distributions represent in certain sense the limiting cases of
very weak TWBs (with mean photon-pair numbers much lower than 1) and stronger
TWBs with their properties developed towards 'the classical limit' (mean
photon-pair numbers in hundreds). Significant differences in their properties,
as discussed below, originate in their joint quasi-distributions $ P_{\rm si} $
of integrated intensities. We recall here that integrated intensities are
introduced in the detection theory \cite{Perina1991} in which the moments of
integrated intensities refer to the normally-ordered photon-number moments. The
quasi-distributions of integrated intensities for a given field-operator
ordering parameter $ s $ are derived from the corresponding quasi-distributions
of field amplitudes defined in the field phase space. The quasi-distributions
of integrated intensities for $ s $-ordered field operators can be determined
from the corresponding joint photon-number distributions $ p_{\rm si} $ along
the formula \cite{Perina1991}
\begin{eqnarray} 
 P_{{\rm si},s}(W_{\rm s},W_{\rm i})&=& \frac{4}{(1-s)^2} \exp\left(-\frac{2(W_{\rm s}+W_{\rm i})}{1-s}\right)
 \nonumber \\
 & & \hspace{-12mm} \times \sum_{n_{\rm s},n_{\rm i} =0}^{\infty}
  \frac{p_{\rm si}(n_{\rm s},n_{\rm i}) }{ n_{\rm s}!\, n_{\rm i}! }
  \left(\frac{s+1}{s-1}\right)^{n_{\rm s}+n_{\rm i}}  \nonumber \\
 & & \hspace{-12mm} \times L_{n_{\rm s}}\left(\frac{4W_{\rm s}}{1-s^2}\right)
   L_{n_{\rm i}}\left(\frac{4W_{\rm s}}{1-s^2}\right)
\label{17}
\end{eqnarray}
in which the symbol $ L_k $ denotes the $ k $-th Laguerre polynomial
\cite{Morse1953}. In the quasi-distribution $ P_{\rm si} $ of the weakest TWB
plotted in Fig.~\ref{fig4}(c), there occur local positive peaks and negative
dips forming the structure with typical rays running from the beginning $
(W_{\rm s},W_{\rm i}) = (0,0) $ and parabolas. On the other hand, the ray
structure with the global positive maximum around the diagonal and negative
'valleys' sandwiched between positive local maxima characterize the
quasi-distribution $ P_{\rm si} $ of the strongest analyzed compound TWB
plotted in Fig.~\ref{fig4}(d). This reflects a well-formed pair-wise character
of the compound TWB with 102 mean photon pairs.

\section{Analysis of effective detection efficiency}

Before we analyze the statistical properties of compound TWBs and their
applications, we have to address first the effective detection efficiencies of
PNRDs used in TWB detection. The effective detection efficiency $ \eta^{\rm
eff} $ of a PNRD is an important parameter when experimental TWBs are analyzed.
It arises from the generalization \cite{Brida2010,PerinaJr2012a} of the Klyshko
method \cite{Klyshko1980} for the determination of absolute detection
efficiency developed originally for individual photon pairs. In fact it gives a
refined effective value of the usual overall detection efficiency that involves
in a real experiment the detector absolute quantum detection efficiency
(estimated at 0.80 for the used APDs and appropriate wavelength) and losses on
the way from the field source to the detector (dominantly caused by fiber
coupling, estimated coupling efficiency below 0.35). This generalization
suggests an appropriate effective detection efficiency $ \eta^{\rm eff}_{\rm s}
$ of the signal detector in the form:
\begin{equation}  
 \eta^{\rm eff}_{\rm s} = \frac{ \langle \Delta c_{\rm s}\Delta c_{\rm i}\rangle }{
  \langle c_{\rm i}\rangle }.
\label{18}
\end{equation}
The effective detection efficiency $ \eta^{\rm eff} $ depends on TWB properties
affected, among others, by the pump-pulse intensity fluctuations and TWB
noises.

We first reveal the limitation of formula (\ref{18}) by considering an intense
genuine TWB described by the photon-number distribution $ p_{\rm si} $ in
Eq.~(\ref{4}). Assuming detection of such TWB by detectors with the detection
efficiencies $ \eta_{\rm s} $ and $ \eta_{\rm i} $ and using the Mandel
detection formula, the correlation function $ \langle \Delta c_{\rm s}\Delta
c_{\rm i}\rangle = \eta_{\rm s} \eta_{\rm i} ( \langle {\cal W}_{\rm p}\rangle
+ \langle (\Delta {\cal W}_{\rm p})^2\rangle ) $ [$ \langle {\cal W}_a\rangle =
M_a B_a $ for $ a = {\rm p,s,i} $, $ \langle (\Delta {\cal W}_{\rm p})^2\rangle
= M_{\rm p}B_{\rm p}^2 $], depends only on the paired component of the genuine
TWB described by the integrated intensity $ {\cal W}_{\rm p} $
\cite{PerinaJr2012a}. On the other hand, the mean idler photocount number $
\langle c_{\rm i}\rangle = \eta_{\rm i} [ \langle {\cal W}_{\rm p}\rangle +
\langle {\cal W}_{\rm i}\rangle ] $ also depends on the noise idler component
with the integrated intensity $ {\cal W}_{\rm i} $. Under these conditions, the
effective detection efficiency $ \eta^{\rm eff}_{\rm s} $ in Eq.~(\ref{18}) is
derived in the form:
\begin{equation}  
 \eta^{\rm eff}_{\rm s} = \eta_{\rm s}
  \frac{ \langle {\cal W}_{\rm p}\rangle + \langle (\Delta {\cal W}_{\rm p})^2\rangle }{
   \langle {\cal W}_{\rm p}\rangle + \langle {\cal W}_{\rm i}\rangle}.
\label{19}
\end{equation}
According to Eq.~(\ref{19}), if a TWB is noiseless [$ \langle {\cal W}_{\rm
i}\rangle = 0 $] and photon-pair statistics is Poissonian [$ \langle (\Delta
{\cal W}_{\rm p})^2\rangle = 0 $] \cite{Perina1991}, we directly have $
\eta^{\rm eff}_{\rm s} = \eta_{\rm s} $. Otherwise, the photon-pair number
fluctuations exceeding the Poissonian ones increase the effective efficiency $
\eta^{\rm eff}_{\rm s} $. The noise acts in the opposed way.

\begin{figure}  
 \includegraphics[width=0.99\hsize]{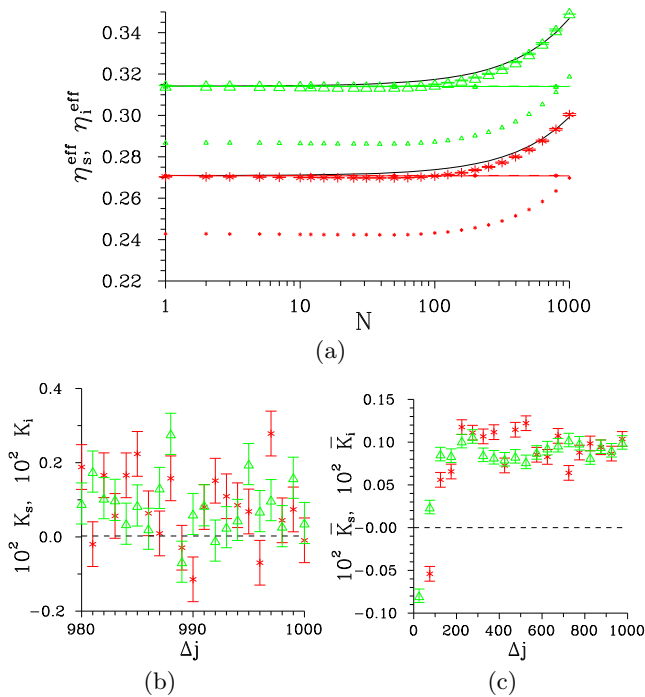}

 \caption{(a) Effective detection efficiencies $ \eta^{\rm eff} $ obtained with (large symbols)
  and without (small symbols) subtraction of the detector dark counts are plotted as
  functions of the number $ N $ of grouped detection windows. (b) Normalized photocount correlation
  functions $ K $ and (c) averaged normalized photocount correlation functions
  $ \bar{K} $ as they depend on the relative distance $ \Delta j $ of detection
  windows; $ \bar{K}_{\Delta j} \equiv \sum_{\Delta k \in \langle \Delta j - \delta j,
  \Delta j + \delta j\rangle} K_{\Delta k} / (2\delta j+1) $. Experimental data are plotted
  as isolated symbols with error bars [in (a) smaller than the plotted symbols] for the signal (red, $ \ast $) and
  idler (green, $ \triangle $) detectors. Solid (dashed)
  curves with symbols in (a) originate in the model of compound (genuine) TWBs detected by APDs ($ N $-pixel PNRDs), the curves nearly
  coincide. Black plain solid curves in (a) are drawn for the model of compound TWBs with
  correlations in intra pump-pulse intensity fluctuations.
  The dashed horizontal lines in (b) and (c) indicate neutral correlations
  $ K = \bar{K} = 0 $; $ \delta j = 24 $.}
\label{fig5}
\end{figure}

In the experiment, the effective signal- and idler-field detection efficiencies
$ \eta^{\rm eff}_{\rm s} $ and $ \eta^{\rm eff}_{\rm i} $ are more-less
constant for the number $ N $ of grouped detection windows up to 100 and then
they gradually increase with the increasing number $ N $ [see
Fig.~\ref{fig5}(a)]. It follows from the curves in Fig.~\ref{fig5}(a) that the
subtraction of the known dark count rates $ D_{\rm s} = 2.8 \times 10^{-3} $
and $ D_{\rm i} = 3.8 \times 10^{-3} $ from the experimental mean photocount
numbers $ \langle c_a\rangle $ in Eq.~(\ref{18}) increases the effective
detection efficiencies $ \eta^{\rm eff}_a $, $ a={\rm s,i} $, by about 2~\%. On
the other hand, the values of the detection efficiencies $ \eta_{\rm s} = 0.282
$ and $ \eta_{\rm i} = 0.330 $ used in the theoretical fit of the experimental
data in Fig.~\ref{fig5}(a) are about 1~\% greater than the measured effective
detection efficiencies $ \eta^{\rm eff} $ for $ N \le 100 $.
This is a consequence of little amount of the noise present in the experimental
compound TWBs. According to the formula in Eq.~(\ref{19}), the theoretical
effective detection efficiencies $ \eta^{\rm eff} $ are independent of the
number $ N $ of grouped detection windows as both expressions in the numerator
and denominator are linearly proportional to the number $ N $ of grouped
detection windows, as documented by the curves in Fig.~\ref{fig5}(a). The
effective detection efficiencies $ \eta^{\rm eff} $ appropriate to $ N $-pixel
PNRDs behave very similarly, as it follows from the curves in
Fig.~\ref{fig5}(a). This behavior originates in the fact that the mean number
of photons impinging on one detection pixel is considerably smaller than one.
However, we note here that for small numbers $ m_{\rm p} $ of paired modes in a
constituting TWB (in one detection window), i.e. when the TWB photon-number
distribution is close to the thermal one, the efficiencies $ \eta^{\rm eff} $
slightly decrease for small numbers $ N $. This effect is described by the term
$ \langle (\Delta {\cal W}_{\rm p})^2\rangle $ in Eq.~(\ref{19}).

The increase of effective detection efficiencies $ \eta^{\rm eff} $ for $ N \ge
100 $ is attributed to correlations in intra pump-pulse intensity fluctuations
already observed in the graphs of the marginal Fano factors $ F_{c,{\rm i}} $
and $ F_{n,{\rm i}} $ in Fig.~\ref{fig3}(b). To quantify these correlations, we
determine the normalized correlation function $ K_{a,\Delta j} $ of photocount
fluctuations $ \Delta c_{a,j} $ and $ \Delta c_{a,j+\Delta j} $ of field $ a $,
$ a={\rm s,i} $, mutually shifted by $ \Delta j $ detection windows for large
number $ N_M $ of measurement repetitions:
\begin{equation}  
 K_{\Delta j} = N_M\frac{ \sum_{j=1}^{N_M} \Delta c_{a,j} \Delta c_{a,j+\Delta j} }{
   [ \sum_{j=1}^{N_M} c_{a,j} ]^2 }.
\label{20}
\end{equation}
When processing the experimental data, the sum over index $ j $ runs over all
detection windows. Random fluctuations of the normalized correlation functions
$ K_{\rm s} $ and $ K_{\rm i} $ for $ \Delta j \in \langle 980, 1000 \rangle $
are documented in Fig.~\ref{fig5}(b). Despite the large relative distance $
\Delta j $ between the detection windows, they tend to exhibit positive
correlations. The normalized correlation functions $ \bar{K}_{\rm s} $ and $
\bar{K}_{\rm i} $ averaged over 49 neighbor indices $ \Delta j $ and plotted in
Fig.~\ref{fig5}(c) clearly express positive correlations for $ \Delta j $
comparable or greater than 100. On the other hand, negative correlations are
observed for $ \Delta j $ smaller than 100. They originate in the electronic
response of the detector in general known as the death-time effect. The
correlations quantified in the graphs of Figs.~\ref{fig5}(b,c) provide suitable
values of parameters of the model for correlations in intra pump-pulse
intensity fluctuations and compound TWB generation developed in Appendix. The
resultant curves for the effective detection efficiencies $ \eta^{\rm eff} $
are drawn in Fig.~\ref{fig5}(a) by plain solid black curves.

\section{Non-classicality of compound twin beams}

The most striking property of TWBs is their non-classicality that originates in
tight correlations between the signal and idler photon numbers
\cite{Jedrkiewicz2004,Haderka2005a,Bondani2007,Blanchet2008,Brida2009a}. As the
marginal signal and idler beams are classical, the TWB non-classicality in fact
reflects the entanglement (quantum correlations) between the signal and idler
beams. We note that this entanglement is different from that usually discussed
for individual photon pairs in quantum superposition states \cite{Mandel1995}.
For compound TWBs, the amount of their overall noise increases linearly with
the number of included constituting TWBs and so the question is to which extent
the non-classicality is preserved for stronger compound TWBs. We show that the
non-classicality of compound TWBs even increases with their intensity, owing to
their increasing photon-pair number.

The non-classicality of TWBs is identified by non-classicality identifies
(NCIs), usually based on the non-classicality inequalities written either in
(integrated) intensity moments \cite{PerinaJr2017a} or containing the
probabilities of photon-number (or photocount) distributions
\cite{PerinaJr2020a}. Here, we restrict our attention to the NCIs based on the
intensity moments. The intensity moments $ \langle W_{\rm s}^k W_{\rm
i}^l\rangle $ are derived from the photon-number moments $ \langle n_{\rm s}^m
n_{\rm i}^j \rangle $ using the following formula
\begin{eqnarray}  
 \langle W_{\rm s}^k W_{\rm i}^l\rangle = \sum_{m=0}^k
   S(k,m) \sum_{j=0}^{l}S(l,j) \langle n_{\rm s}^m n_{\rm i}^j \rangle
\label{21}
\end{eqnarray}
that includes the Stirling numbers $ S $ of the first kind
\cite{Gradshtein2000}. We note that the relations in Eq.~(\ref{21}) between the
normally-ordered photon-number moments (referred as intensity moments) and the
usual photon-number moments originate in the canonical commutation relations.
We also address the non-classicality of the directly detected photocount
distributions $ f_{\rm si}^{\rm c} $. To do this, we pretend as if the analyzed
photocount distributions $ f_{\rm si}^{\rm c} $ were obtained by ideal
detectors and determine the associated intensity moments $ \langle W_{\rm s}^k
W_{\rm i}^l\rangle_c $ from the photocount moments $ \langle c_{\rm s}^m c_{\rm
i}^j \rangle $ along the formula (\ref{21}).

The intensity NCIs for TWBs were comprehensively analyzed in
\cite{PerinaJr2017a}. Our motivation here is twofold. First, to study the
higher-order non-classicality indicated by higher-order intensity moments for
the compound TWBs with the intensities varying over three orders in magnitude.
Second, to identify the best performing NCIs, especially for greater numbers $
N $ of grouped detection windows. In our analysis, the following NCIs
containing the intensity moments up to the fifth order were found the best
\cite{PerinaJr2017a}:
\begin{eqnarray}  
 E_{001} &\equiv& \langle W_{\rm s}^2\rangle + \langle W_{\rm i}^2 \rangle - 2\langle W_{\rm s}
  W_{\rm i}\rangle <0, \label{22} \\
 E_{101} &\equiv& \langle W_{\rm s}^3\rangle + \langle W_{\rm s}W_{\rm i}^2\rangle - 2\langle W_{\rm s}^2
  W_{\rm i}\rangle <0, \label{23} \\
 E_{111} &\equiv& \langle W_{\rm s}^3W_{\rm i}\rangle + \langle W_{\rm s}W_{\rm i}^3\rangle -
  2\langle W_{\rm s}^2W_{\rm i}^2\rangle <0, \label{24} \\
 E_{211} &\equiv& \langle W_{\rm s}^4W_{\rm i}\rangle + \langle W_{\rm s}^2W_{\rm i}^3\rangle -
  2\langle W_{\rm s}^3W_{\rm i}^2\rangle <0. \label{25}
\end{eqnarray}
Only the following two NCIs derived from the matrix approach were capable to
indicate the non-classicality of the directly measured photocount distributions
$ f_{\rm si}^{\rm c} $ for greater numbers $ N $ of grouped detection windows:
\begin{eqnarray}  
 M_{1001} &\equiv& \langle W_{\rm s}^{2}\rangle\langle W_{\rm i}^{2}\rangle
  - \langle W_{\rm s}W_{\rm i}\rangle^2 < 0,
\label{26}  \\
 M_{001001} &\equiv& \langle W_{\rm s}^{2}\rangle\langle W_{\rm i}^{2}\rangle
   + 2\langle W_{\rm s}W_{\rm i}\rangle\langle W_{\rm s}\rangle\langle W_{\rm i}\rangle
   - \langle W_{\rm s}W_{\rm i}\rangle^2 \nonumber \\
  & & \mbox{} - \langle W_{\rm s}^2\rangle\langle W_{\rm i}\rangle^2
   - \langle W_{\rm s}\rangle^2\langle W_{\rm i}^2\rangle < 0.
   \label{27}
\end{eqnarray}

The above NCIs can even be used for quantifying the non-classicality when the
corresponding Lee non-classicality depths (NCD) \cite{Lee1991} are determined.
To do this, we first determine the intensity moments $ \langle W_{\rm s}^k
W_{\rm i}^l \rangle_s $ related to a general $ s $ ordering of the field
operators from the normally-ordered intensity moments $ \langle W_{\rm s}^k
W_{\rm i}^l \rangle $ using the coefficients of the Laguerre polynomials $ L_k
$ ~\cite{Perina1991}:
\begin{equation}   
 \langle W_{\rm s}^k W_{\rm i}^l \rangle_s = \left(\frac{1-s}{2}\right)^{k+l} \left\langle
  {\rm L}_k \left(\frac{2W_{\rm s}}{s-1}\right) {\rm L}_l
  \left(\frac{2W_{\rm i}}{s-1}\right)\right\rangle.
\label{28}
\end{equation}
Then we apply the above NCIs written directly for the $ s $-ordered intensity
moments $ \langle W_{\rm s}^k W_{\rm i}^l \rangle_s $. The non-classicality
indicated by the NCIs is gradually suppressed with the decreasing value of the
ordering parameter $ s $ due to the increasing detection noise related to the
operator ordering. Using this we look for the threshold value $ s_{\rm th} $ at
which a given NCI nullifies. It gives the border between the quantum and the
classical behavior. The corresponding NCD $ \tau $ is derived as follows
\cite{Lee1991}
\begin{equation}    
 \tau = \frac{1-s_{\rm th} }{2}.
\label{29}
\end{equation}
It gives the least mean number of thermal photons of a noisy field that, when
superimposed on the analyzed field, guarantees the suppression of
non-classicality of the analyzed field. It holds that $ 0 < \tau \le 1/2 $ for
any non-classical 2D Gaussian field.

\begin{figure}  
 \includegraphics[width=0.99\hsize]{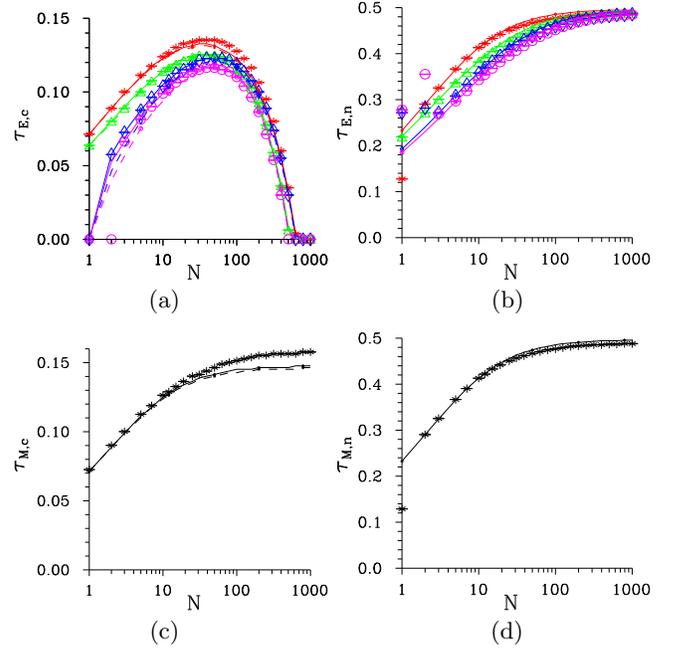}

 \caption{Non-classicality depths for (a,c) the joint photocount ($ \tau_{c} $) and (b,d)
 the joint photon-number ($ \tau_{n} $) distributions as they depend on the number $ N $ of
 grouped detection windows. The NCIs $ E_{001} $ (red, $ \ast $), $ E_{101} $ (green, $ \triangle $),
 $ E_{111} $ (blue, $ \diamond $), and $ E_{211} $ (magenta, $ \circ $) are used in
 (a,b), the NCI $ M_{1001} $ ($ \ast $) is drawn in (c,d). Experimental data are plotted by isolated
 symbols, their error bars are smaller than the plotted symbols. Solid (dashed) curves
 originate in the model of compound (genuine) TWBs detected by APDs ($ N $-pixel PNRDs);
 the solid and dashed curves nearly coincide.}
\label{fig6}
\end{figure}

Whereas the NCIs $ E $ defined in Eqs.~(\ref{22})---(\ref{25}) identify the
non-classicality for any reconstructed compound TWB, they fail to resolve the
non-classicality directly contained in the corresponding photocount
distributions $ f_{\rm si}^c $ belonging to stronger compound TWBs (for $ N \ge
800 $), as documented by the curves of NCDs $ \tau_{E,c} $ and $ \tau_{E,n} $
in Figs.~\ref{fig6}(a,b). This is a consequence of the noise present in TWB
photocount distributions $ f_{\rm si}^c $ whose level increases linearly with
the number $ N $ of grouped detection windows. The reached values of NCDs $
\tau_{E,c} $ and $ \tau_{E,n} $ slightly decrease with the increasing order of
the involved intensity moments. Nevertheless, the NCDs $ \tau_{E,n} $ for the
reconstructed compound TWBs tend to approach the values close to the maximal
allowed value 1/2 for large numbers $ N $ of grouped detection windows. On the
other hand, the maximal values of NCDs $ \tau_{E,c} $ for the photocount
distributions $ f_{\rm si}^c $ close to 0.14 are reached for $ N $ around 50.

The NCIs $ M $ written in Eqs.~(\ref{26}) and (\ref{27}) reveal the
non-classicality also for the photocount distributions $ f_{\rm si}^c $ of
strong composed TWBs, as shown in Fig.~\ref{fig6}(c) for the NCI $ M_{1001} $.
Moreover, the obtained values of NCDs $ \tau_{M,c} $ increase with the number $
N $ of grouped detection windows, similarly as those of NCDs $ \tau_{M,n} $
appropriate for the reconstructed compound TWBs [Fig.~\ref{fig6}(d)]. We note
that the values of the NCDs $ \tau_{M} $ belonging to the NCIs in
Eqs.~(\ref{26}) and (\ref{27}) are for the analyzed compound TWBs very close to
each other independently of the number $ N $. As the NCI $ M_{001001} $ in
Eq.~(\ref{27}) has a more complex structure than the NCI $ M_{1001} $
in~(\ref{26}), it is more prone to the experimental errors. That is why we
prefer to use the NCI $ M_{1001} $ for identification of the non-classicality
of compound TWBs of arbitrary intensity. Both theoretical models for TWBs and
their detection give good predictions for the NCDs $ \tau $ derived from
photon-number and photocount distributions, as it is shown in Fig.~\ref{fig6}:
The model of compound TWBs detected by APDs predicts slightly larger values of
the NCDs $ \tau $ for photocount distributions compared to the model of genuine
TWBs detected by $ N $-pixel PNRDs (stronger pile-up effect).

\section{Sub-Poissonian light generation}

Compound TWBs may also be used for efficient generation of sub-Poissonian light
via post-selection. Post-selection by PNRDs applied to TWBs belongs, together
with the cavity quantum electrodynamics schemes \cite{Raimond2001}, to the most
efficient methods of sub-Poissonian-light generation with greater photon
numbers, both in cw \cite{Rarity1987,Laurat2003,Zou2006} and pulsed regimes
\cite{Bondani2007,PerinaJr2013b,Lamperti2014,Iskhakov2016,Harder2016}. However,
for realistic PNRDs, the method gradually loses its potential with the
increasing post-selecting photocount number which limits the maximal
intensities of the generated sub-Poissonian fields. Here, we address this
limitation for compound TWBs and APDs.

An idler field conditioned by the detection of $ c_{\rm s} $ signal photocounts
is experimentally characterized by the conditional photocount histogram $
f_{c,\rm i}^{\rm c}(c_{\rm i};c_{\rm s}) $ derived from the joint signal-idler
photocount histogram $ f_{\rm si}^{\rm c}(c_{\rm s},c_{\rm i}) $ by appropriate
normalization:
\begin{equation}   
 f_{c,\rm i}^{\rm c}(c_{\rm i};c_{\rm s}) = \frac{ f_{\rm si}^{\rm c}(c_{\rm s},c_{\rm i}) }{
  \sum_{c'_{\rm i}=0}^{\infty} f_{\rm si}^{\rm c}(c_{\rm s},c'_{\rm i}) }.
\label{30}
\end{equation}
The corresponding conditional idler photon-number distributions $ p_{c,\rm
i}^{\rm c}(n_{\rm i};c_{\rm s}) $ can then be obtained using the ML approach,
similarly as in the case of TWBs. The looked-for distribution $ p_{c,\rm
i}^{\rm c}(n_{\rm i};c_{\rm s},N) $ obtained from a compound TWB formed by $ N
$ constituting TWBs (detected in $ N $ detection windows of APDs) is found as a
steady state of the following iteration procedure ($ j $ stands for the
iteration index):
\begin{eqnarray}   
 p^{{\rm c}(j+1)}_{c,\rm i}(n_{\rm i};c_{\rm s},N) &=& \sum_{c_{\rm i}=0}^{\infty}
  F^{(j)}_{\rm i}(c_{\rm i};c_{\rm s},N) T_{\rm i}(c_{\rm i},n_{\rm i};N),
\label{31} \\
 F^{(j)}_{\rm i}(c_{\rm i};c_{\rm s},N) &=& f^{\rm c}_{c,\rm i}(c_{\rm i};c_{\rm s},N) \Biggl[
  \sum_{n'_{\rm i}=0}^{\infty} T_{\rm i}(c_{\rm i},n'_{\rm i};N)   \nonumber \\
 & & \hspace{-20mm} \times p^{{\rm c}(j)}_{c,\rm i}(n'_{\rm i};c_{\rm s},N) \Bigr]^{-1},
  \hspace{2mm} j=0,1,\ldots \; . \nonumber
\end{eqnarray}

Sub-Poissonianity is quantified by the Fano factor $ F $ defined in
Eq.~(\ref{14}) that naturally depends on the conditioning signal photocount
number $ c_{\rm s} $. Detailed inspection \cite{PerinaJr2013b,PerinaJr2017c}
reveals its following behavior that reflects the limited detection efficiency
and the presence of dark counts in the post-selecting detector: With the
increasing conditioning signal photocount number $ c_{\rm s} $, the values of
Fano factor $ F $ decrease naturally first, they find their minimum and then
they increase at the greater signal photocount numbers $ c_{\rm s} $ that have
very low probability of detection. As we are interested in the potential of the
post-selection scheme to generate the most nonclassical states as possible, we
further analyze the values of the Fano factor in this minimum.

\begin{figure}  
 \includegraphics[width=0.99\hsize]{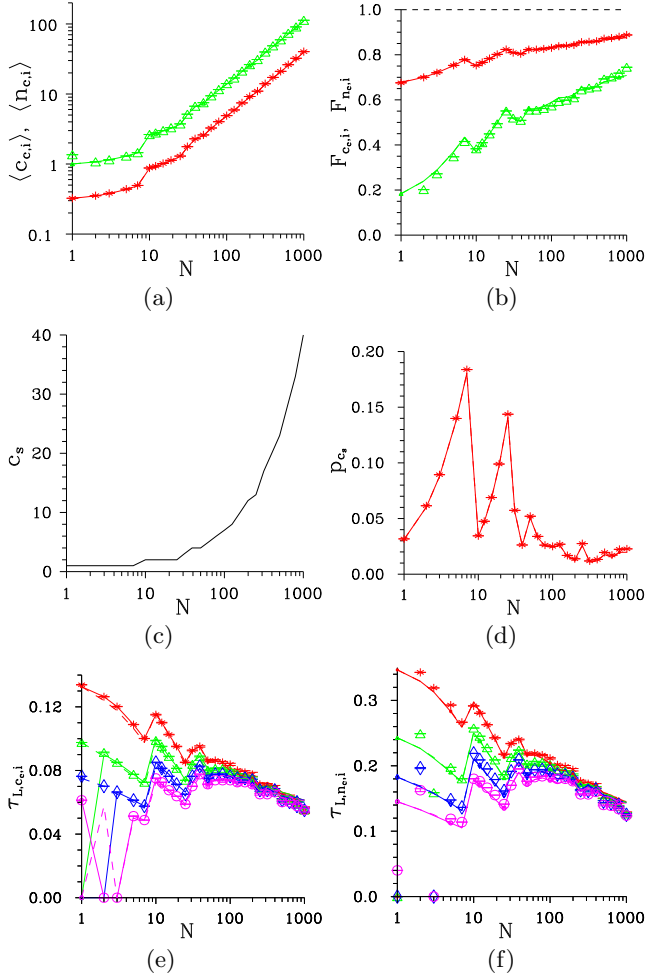}

 \caption{(a) Mean number of idler photons $ \langle n_{c,\rm i}\rangle $ (photocounts $ \langle c_{c, \rm i}\rangle $), (b)
  Fano factor $ F_{n_c,{\rm i}} $ ( $ F_{c_c,{\rm i}} $), (c) optimal conditioning signal photocount number $ c_{\rm s}$
  and (d) its probability $ p_{c_{\rm s}} $ of detection and (e)
  [(f)] photocount [photon] number NCDs $ \tau_{L,c_c,{\rm i}} $ [$ \tau_{L,n_c,{\rm i}} $] for NCIs $ L_{11}^1 $ (red, $ \ast $), $
  L_{21}^1 $ (green, $ \triangle $), $ L_{31}^1 $ (blue, $ \diamond $), and $ L_{41}^1 $ (magenta, $ \circ $)
  belonging to conditional idler fields $ p_{c,{\rm i}}^{\rm c} $ generated after
  registering the $ c_{\rm s} $ optimal signal photocounts indicated in (c) and depending on the number $ N $ of grouped detection windows.
  In (a), (b) and (d), isolated symbols with error bars (smaller than the plotted symbols) are drawn for the experimental photocount
  histograms (red $ \ast $) and fields reconstructed by the ML approach (green $ \triangle $).
  Experimental data plotted in (e) and (f) by isolated symbols have error bars smaller than the used symbols.
  Solid (dashed) curves arise in the model of compound (genuine) TWBs detected by APDs ($ N $-pixel PNRDs).
  In (a), (b), (d), (e) and (f), the solid and dashed curves nearly coincide.
  The horizontal plain dashed line in (b) indicates the non-classicality border $ F = 1 $.}
\label{fig7}
\end{figure}

With the increasing number $ N $ of grouped detection windows, the intensity of
compound TWBs increases, the optimal conditioning signal photocount number $
c_{\rm s} $ increases [Fig.~\ref{fig7}(c)] and, as a consequence, the mean
idler photocount ($ \langle c_{c,\rm i}\rangle $) as well as photon ($ \langle
n_{c,\rm i}\rangle $) numbers of the post-selected fields monotonously
increase, as quantified in Fig.~\ref{fig7}(a). Unfortunately, the accompanying
photocount and photon-number Fano factors $ F_{c_c,{\rm i}} $ and $ F_{n_c,{\rm
i}} $, respectively,  also increase with the increasing number $ N $ of grouped
detection windows [see Fig.~\ref{fig7}(b)]. This increase originates in the
increasing level of the overall noise in the post-selecting signal detector
with the increasing number $ N $ of grouped detection windows. This noise has
strong detrimental influence to the performance of the post-selection scheme.
Whereas the idler field post-selected by detecting in just one detection window
has around 1 photon and the Fano factor around 0.2, the idler field
post-selected by the detection in 1000 grouped detection windows contains
around 100 photons, but its Fano factor is only around 0.7. We note that the
teeth in the curves of Fano factors $ F $ in Fig.~\ref{fig7}(b) are caused by
discrete values of the optimal conditioning signal photocount numbers $ c_{\rm
s} $. The analyzed scheme for the generation of sub-Poissonian light is
probabilistic and the probability $ p_{c_{\rm s}} $ of success in general
decreases with the number $ N $ of grouped detection windows. However, as it
follows from the curve in Fig.~\ref{fig7}(d), the probability $ p_{c_{\rm s}} $
of success is in percents even for the strongest analyzed compound TWBs.

Similarly as in the case of compound TWBs, we address the non-classicality of
the conditional idler fields $ p_{c,\rm i}^{\rm c} $ using the NCIs based on
intensity moments up to the fifth order. Relying on the results of
Ref.~\cite{PerinaJr2020}, we apply the following fundamental NCIs:
\begin{equation} 
 L_{k1}^1 = \langle W^{k+1}\rangle - \langle W^k\rangle \langle W\rangle < 0,
  \hspace{5mm} k=1,\ldots,4.
\label{32}
\end{equation}
We note that the commonly used NCIs $ \langle W^k\rangle - \langle W\rangle^k <
0 $, $ k=2,\ldots,5 $, are derived with the help of the NCIs in Eq.~(\ref{32}).
We can see in Figs.~\ref{fig7}(e,f) that the photocount NCDs $ \tau_{L,c_c,{\rm
i}} $ as well as the photon NCDs $ \tau_{L,n_c,{\rm i}} $ gradually decrease
with the increasing number $ N $ of grouped detection windows, i.e. with the
increasing intensity of the post-selected idler fields. Whereas the attained
values of the NCDs $ \tau_{L,{\rm i}} $ considerably drop down with the order
of intensity moments involved in the NCIs for smaller numbers $ N $, the values
of the NCDs $ \tau_{L,{\rm i}} $ are close to each other for greater numbers $
N $. In Figs.~\ref{fig7}(e,f), the reached values of the photon NCDs $
\tau_{L,n_c,{\rm i}} $ are about three times greater than the values of the
corresponding  photocount NCDs $ \tau_{L,c_c,{\rm i}} $, in agreement with the
detection efficiency $ \eta_{\rm i} = 0.33 $.

We note that the models of compound TWBs detected by APDs [see Eq.~(\ref{11})]
and genuine TWBs detected by $ N $-pixel PNRDs give, apart from small numbers $
N $, very similar predictions of the quantities characterizing the
post-selected sub-Poissonian fields that are in good agreement with the
experimental results, as documented in the graphs of Fig.~\ref{fig7}.

\section{Sub-shot-noise measurement of absorption}

As we have seen in the previous section, post-selection with compound TWBs
gives compound sub-Poissonian fields with the reduced photon-number
fluctuations. When applied to the measurement of absorption coefficients
\cite{Jakeman1986,Genovese2016,Whittaker2017,Losero2018,SabinesChesterkind2019}
they provide the sub-shot-noise precision, i.e. the precision better than the
classical optimal limit reached by the coherent states with their Poissonian
photon-number statistics. Below we show that the measurement precision
increases with the increasing intensity of a compound TWB.

Determination of mean values $ \langle c \rangle $ of photocounts is in the
essence of the measurement of absorption coefficients. The mean photocount
numbers $ \langle c \rangle $ are measured for the fields in front and beyond
the sample and their ratio gives the looked-for absorption coefficient. The
precision (uncertainty) of such measurement is quantified by the relative error
$ \delta c $ given by
\begin{equation}   
 \delta c = \frac{ \sqrt{\langle (\Delta c)^2\rangle_{\rm m}} }{ \langle c\rangle_{\rm m}}
\label{33},
\end{equation}
and the symbol $ \langle \;\rangle_{\rm m} $ denotes averaging over the
obtained finite experimental data set. If a coherent state with mean photocount
number $ \langle c\rangle_{\rm cl} $ and Poissonian photocount statistics [$
\langle (\Delta c)^2\rangle_{\rm cl} = \langle c\rangle_{\rm cl} $] is applied
$ N_M $ times, the corresponding relative error $ \delta c_{\rm cl} $ takes the
form:
\begin{equation}   
 \delta c_{\rm cl} = \frac{1}{ \sqrt{ \langle c\rangle_{\rm cl} N_M }}.
\label{34}
\end{equation}
It serves as a reference in the definition of the normalized relative error $
\delta_r c $:
\begin{equation}   
 \delta_r c = \frac{ \delta c }{ \delta c_{\rm cl} }.
\label{35}
\end{equation}
Sub-shot-noise measurements are characterized by $ \delta_r c < 1 $.

To demonstrate the ability of compound TWBs to overcome the classical limit of
Eq.~(\ref{34}), we analyze in parallel two data sets. Speaking about the idler
detector, we consider the sequence of idler detections in all detection windows
as a reference. The second sequence contains the idler detections only in the
detection windows in which a signal photocount was registered. If we sum the
photocount numbers $ c_{\rm i} $ in $ N $ subsequent detection windows in the
first reference sequence, we arrive at the photocount histogram characterizing
the marginal idler field of the compound TWB formed by $ N $ constituting TWBs.
On the other hand, the summation of the conditioned photocount numbers $ c_{c,
{\rm i}} $ over $ N $ subsequent detection windows in the second sequence
provides us the photocount histogram of the idler field conditioned by the
detection of $ N $ signal photocounts in $ N $ grouped detection windows
(forming a compound TWB). This field is sub-Poissonian and overcomes the
classical limit of Eq.~(\ref{34}).

\begin{figure}  
 \includegraphics[width=0.8\hsize]{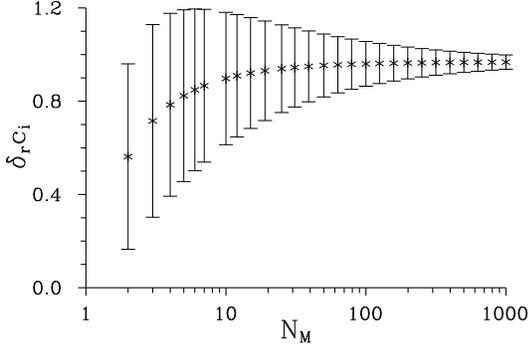}

 \caption{Normalized relative error $ \delta_r c_{\rm i} $ of the idler  photocount number for
  $ N = 200 $ grouped detection windows as it depends on the number $ N_M $ of
  measurement repetitions. The plotted relative experimental errors of $ \delta_r c_{\rm i} $
  are given as $ 1/\sqrt{N_M} $.}
\label{fig8}
\end{figure}

To experimentally quantify the dependence of measurement precision on the
number $ N_M $ of measurement repetitions, we determine the following
experimental relative error:
\begin{equation}   
 \delta c(N_M) =  \left\langle  \frac{ \sqrt{ \sum_{j=1}^{N_M} c_j^2 /N_M -
   \left( \sum_{j=1}^{N_M} c_j /N_M \right)^2 }}{ \sum_{j=1}^{N_M} c_j /N_M }
   \right\rangle_{\rm m}.
\label{36}
\end{equation}
However, we have to consider sufficiently large numbers $ N_M $ of measurement
repetitions to arrive at reliable (acceptable) values of the relative errors $
\delta c $. We illustrate this behavior in Fig.~\ref{fig8}, where the
normalized relative error $ \delta_r c_{\rm i} $ of the idler photocount
numbers $ c_{\rm i} $ summed over $ N=200 $ detection windows is plotted. The
values of the normalized relative error $ \delta_r c_{\rm i} $ for the number $
N_M $ of measurement repetitions smaller than 10 are, though with large
absolute experimental errors, artificially smaller than the correct value found
for the asymptotically large numbers $ N_M $. The data drawn in Fig.~\ref{fig8}
suggest that the numbers $ N_M $ of measurement repetitions greater than 100
provide reliable values of the mean photocount number $ \langle c_{\rm
i}\rangle $ for the analyzed idler field ($\langle c_{\rm i}\rangle = 7.463\pm
0.002 $).

\begin{figure}  
 \includegraphics[width=0.99\hsize]{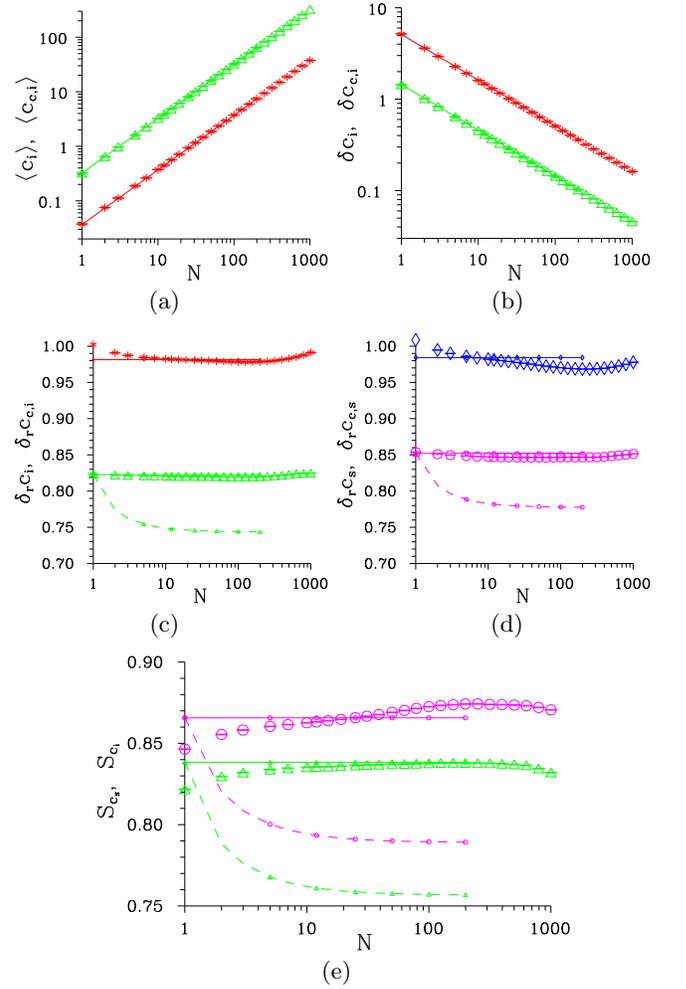}

 \caption{(a) Mean numbers of idler photocounts $ \langle c_{\rm i}\rangle $ and
  conditioned idler photocounts $ \langle c_{c, \rm i}\rangle $ and (b) the
  corresponding relative errors $ \delta c_{\rm i} $ and $ \delta c_{c,\rm i}
  $, (c) [(d)] normalized relative errors $ \delta_r c_{\rm i} $ [$ \delta_r c_{\rm s} $]
  and $ \delta_r c_{c,{\rm i}} $ [$ \delta_r c_{c,{\rm s}} $]
  and (e) ratios $ S_{c_{\rm s}} $ and $ S_{c_{\rm i}} $ of normalized relative errors as they depend on the number
  $ N $ of grouped detection windows. Isolated symbols with error bars (smaller than the plotted symbols)
  originate in the
  experiment with signal [idler] photocounts (blue $ \diamond $ [red $ \ast $])
  and signal [idler] conditioned photocounts (magenta $ \circ $ [green $ \triangle
  $]). Solid (dashed curves) are given by the model of compound (genuine) TWBs
  detected by APDs ($ N $-pixel PNDRs); $ N_M = 500 $.}
\label{fig9}
\end{figure}

According to Eq.~(\ref{34}), the relative error $ \delta c $ decreases with the
increasing field intensity, i.e. with the increasing mean photocount number $
\langle c\rangle $. Working with the compound TWBs, the mean photocount numbers
$ \langle c\rangle $ increase linearly with the number $ N $ of grouped
detection windows. These dependencies are drawn in Figs.~\ref{fig9}(a,b) for
the idler photocount numbers $ c_{\rm i} $ (from the reference sequence) and
conditioned idler photocount numbers $ c_{c,{\rm i}} $. For a given number $ N
$ of grouped detection windows, the mean conditioned photocount numbers $
c_{c,{\rm i}} $ are about ten times greater than the mean photocount numbers $
c_{\rm i} $ of the reference sequence [see Fig.~\ref{fig9}(a)], which is given
by the parameters of the post-selection scheme. On the other hand, the relative
errors $ \delta c_{c,{\rm i}} $ belonging to the conditioned photocounts are
more than $ \sqrt{10} $ times smaller than the relative errors $ \delta c_{\rm
i} $ of the photocounts from the reference sequence [see Fig.~\ref{fig9}(b)].

The reduction of the relative error $ \delta c_{c,{\rm i}} $ of the conditioned
photocounts with respect to the relative error $ \delta c_{\rm i} $ of the
reference sequence by more than $ \sqrt{ \langle c_{c,{\rm i}}\rangle /\langle
c_{\rm i}\rangle } $ is caused by the sub-Poissonian character of the
conditioned idler photocount fields. To separate the effect of field intensity
on the relative errors, we plot in Figs.~\ref{fig9}(c,d) the idler normalized
relative errors $ \delta_r c_{\rm i} $ and $ \delta_r c_{c,{\rm i}} $ together
with their signal counterparts $ \delta_r c_{\rm s} $ and $ \delta_r c_{c,{\rm
s}} $. The values of normalized relative errors $ \delta_r c_{\rm s} $ and $
\delta_r c_{\rm i} $ for photocounts of the reference sequence at the signal
and idler detectors, respectively, are slightly below one. The multi-mode
thermal character of the detected marginal signal and idler fields in the
reference sequence suggests the values of these normalized relative errors
slightly above one, but the pile-up effect in the detection in individual
detection windows leads to the values slightly smaller than one. On the other
hand, the normalized relative errors $ \delta_r c_{c,{\rm s}} \approx 0.82 $
and $ \delta_r c_{c,{\rm i}} \approx 0.85 $ of the conditioned photocounts
express the increased measurement precision due to the sub-Poissonian character
of the conditioned photocount fields.

To quantify the improvement in the measurement precision due to the application
of such compound sub-Poissonian fields, we draw in Fig.~\ref{fig9}(e) the
ratios $ S_{c_{\rm s}} $ and $ S_{c_{\rm i}} $ of the conditioned photocount
normalized relative errors and their counterparts in the reference signal and
idler sequences:
\begin{equation}    
 S_{c_a} = \frac{ \delta_r c_{c,a} }{ \delta_r c_a }, \hspace{5mm} a={\rm s,i} .
\label{37}
\end{equation}
In our experiment and according to the curves in Fig.~\ref{fig9}(e), the
improvement in the measurement precision of absorption coefficient lies in the
range from 12~\% to 17~\%. Better improvement is observed at the idler detector
as the idler detection efficiency $ \eta_{\rm i} $ is greater than the signal
detection efficiency $ \eta_{\rm s} $. The ratios $ S_{c_{\rm s}} $ and $
S_{c_{\rm i}} $ slightly increase with the number $ N $ of grouped detection
windows. This is caused by the increasing level of the noise with the
increasing number $ N $ of grouped detection windows caused by the correlations
in intra pump-pulse intensity fluctuations already discussed in relation to the
Fano factors of the marginal fields and effective detection efficiencies. We
note that slight decrease of the values of the ratios $ S_{c_{\rm s}} $ and $
S_{c_{\rm i}} $ close to $ N = 1000 $ in Fig.~\ref{fig9}(e) is caused by the
fact that we average over $ N_M = 500 $ measurement repetitions for technical
reasons ($ N_M < N $).

In the graphs of Fig.~\ref{fig9}, we also plot the curves arising in the models
of compound TWBs detected by APDs (solid curves) and genuine TWBs detected by $
N $-pixel PNRDs (dashed curves). The quantities for photocount fields in the
reference sequence are directly derived from the signal and idler marginal
fields of compound TWBs formed by $ N $ constituting TWBs. On the other hand,
the quantities for conditioned photocount fields originate in the model of
sub-Poissonian fields obtained from compound TWBs formed by $ N $ constituting
TWBs that are post-selected by registering $ N $ photocounts in the
complementary field. According to the curves in Figs.~\ref{fig9}(c,d,e), the
use of genuine TWBs with $ N $-pixel PNRDs instead of the analyzed compound
TWBs detected in $ N $ APDs detection windows improves the measurement
precision by about 50~\% for $ N > 10 $ (signal detector: $13 \% \rightarrow \;
21 \% $, idler detector: $ 16 \% \rightarrow \; 25 \% $). This partial
degradation in the measurement precision is caused by the structure of compound
TWBs that, however, allows their simple detection by APDs.

We note that we have used the shot-noise limits of the applied detectors with
limited detection efficiencies as a reference and beat this limit by the
generated compound sub-Poissonian fields. If the detectors with detection
efficiencies close to 1 are used, the analyzed experimental scheme would allow
to beat the ultimate shot-noise limit set for an ideal detector. This has been
reached, e.g., in \cite{Losero2018}.

\section{Conclusions}

We have suggested a scheme for substituting genuine multi-mode twin beams by
compound twin beams. The compound twin beams are composed of constituting twin
beams that replace the fields in individual spatio-spectral modes of the
genuine twin beams. The constituting twin beams are sufficiently weak so that
they allow for the detection by single-photon sensitive on/off detectors. As a
consequence the photocount statistics of compound twin beams are revealed
without the need of photon-number-resolving detectors.

We have experimentally analyzed the properties of compound twin beams
containing up to hundreds of photon pairs, side by side with the appropriate
theoretical models. The determined marginal Fano factors, noise-reduction
parameters and covariances of photon numbers revealed close similarity with the
genuine twin beams. We have determined the effective detection efficiencies for
the compound twin beams.

Using experimental compound twin beams with intensities varying over three
orders in magnitude we have identified the non-classicality identifiers
suitable for stronger twin beams and, by determining the corresponding
non-classicality depths, we confirmed the highly nonclassical properties of
stronger compound twin beams.

Using compound twin beams and post-selection based on the photocount
measurement in one beam, we have experimentally generated sub-Poissonian fields
with intensities up to one hundred of photons and systematically studied their
properties including non-classicality quantification. However, we have
experimentally demonstrated that the real post-selection scheme loses its
efficiency with the increasing field intensity.

We have experimentally demonstrated the ability of compound twin beams to
perform sub-shot-noise measurements of absorption. Though the compound twin
beams are less convenient than the genuine twin beams in this measurement, the
use of single-photon sensitive on/off detectors in their detection represents a
huge advantage, both from the point of view of their operation and high
absolute detection efficiencies. Sub-shot-noise imaging with compound twin
beams is especially promising for imaging of biological and other samples prone
to light illumination. Broad spectra of twin beams also allow for spectroscopy
measurements with sub-shot-noise precision.

\acknowledgments The authors thank GA \v{C}R project No.~18-22102S. They also
acknowledge the support from M\v{S}MT \v{C}R (project
No.~CZ.1.05/2.1.00/19.0377).

\appendix
\section{Model for correlated intra pump-pulse intensity fluctuations and compound twin-beam generation}

The comparison of the measured effective detection efficiencies $ \eta^{\rm
eff} $ and the marginal photocount Fano factors $ F_{c} $ drawn in
Figs.~\ref{fig5}(a) and \ref{fig3}(b), respectively, with their theoretical
counterparts shows that there occurs an additional noise that affects the
compound TWBs detected in more than 100 detection windows. This additional
noise is linearly proportional to the number $ N $ of detection windows for $ N
\ge 100 $. As it affects the measured effective detection efficiencies $
\eta^{\rm eff} $, it has to be of the 'paired' origin. We attribute this noise
to the correlations of the pump-pulse intensity fluctuations among subsequent
pulses. These pump-pulse intensity fluctuations are transferred in the process
of spontaneous parametric down-conversion into the fluctuations of integrated
intensity of the paired component of a TWB (the photon-pair number). This then
primarily affects the quantities directly depending on the strength of the
signal and idler field correlations (the detection efficiencies $ \eta^{\rm
eff} $) and secondarily the quantities depending solely on the signal or the
idler field via the accompanying marginal noises.

To account for such correlations in pump-pulse intensity fluctuations in the
first model of compound TWBs detected by APDs, we consider the following
classical statistical Gaussian model for additional correlated intensity
fluctuations $ w_{{\rm p},j} $ and $ w_{{\rm p},k} $ of the TWB paired
components in detection windows $ j $ and $ k $:
\begin{eqnarray}   
 & \langle w_{{\rm p},j} \rangle_w = 0; & \nonumber \\
 & \langle w_{{\rm p},j} w_{{\rm p},k} \rangle_w = ( 1-\delta_{jk}) K \langle {\cal W}_{\rm
  p}^w \rangle^2 . &
\label{A1}
\end{eqnarray}
In writing Eq.~(\ref{A1}), we assume that the strength of the additional
intensity fluctuations $ w_{{\rm p},j} $ is linearly proportional to the mean
intensity $ \langle {\cal W}_{\rm p}^w \rangle $ of the TWB paired component
belonging to one detection window. Provided that the detector dark-count rates
can be neglected, the value of the phenomenological constant $ K $ introduced
in the second relation of Eq.~(\ref{A1}) can be estimated by the averaged
values of coefficients $ \bar{K}_{\rm s} $ and $ \bar{K}_{\rm i} $ drawn in
Fig.~\ref{fig5}(c).

The integrated intensity $ {\cal W}_{\rm p}^{\rm all} $ of the paired component
of a compound TWB originating in $ N $ grouped detection windows and involving
the additional intensity fluctuations is determined as $ {\cal W}_{\rm p}^{\rm
all} \equiv {\cal W}_{\rm p} + \sum_{j=1}^{N} w_{{\rm p},j} $. Its moments with
respect to the additional pump-pulse intensity fluctuations take the form:
\begin{eqnarray}   
 \langle {\cal W}_{\rm p}^{\rm all} \rangle_w &=& \langle {\cal W}_{\rm p}
  \rangle, \nonumber \\
 \langle \left[{\cal W}_{\rm p}^{\rm all}\right]^2 \rangle_w &=& \langle
  {\cal W}_{\rm p}^2 \rangle + K N(N-1) \langle {\cal W}_{\rm p}^w \rangle^2.
\label{A2}
\end{eqnarray}

The original formula (\ref{19}) for the signal effective detection efficiency $
\eta^{\rm eff}_{\rm s} $ is then modified into the form
\begin{equation}  
 \eta^{\rm eff}_{{\rm s}} = \eta_{\rm s}
  \frac{ \langle {\cal W}_{\rm p}\rangle + \langle (\Delta {\cal W}_{\rm p})^2\rangle +
   K N(N-1) \langle {\cal W}_{\rm p}^w \rangle^2 }{
   \langle {\cal W}_{\rm p}\rangle + \langle {\cal W}_{\rm i}\rangle}.
\label{A3}
\end{equation}
Similarly, the signal photocount Fano factor $ F_{c,{\rm s}} $ defined in
Eq.~(\ref{14}) is expressed in this model as follows:
\begin{eqnarray}  
 F_{c,{\rm s}} &=& 1 + \eta_{\rm s}
  \Bigl[ \langle (\Delta {\cal W}_{\rm p})^2\rangle + \langle (\Delta {\cal W}_{\rm s})^2\rangle  \nonumber \\
 & & + K N(N-1) \langle {\cal W}_{\rm p}^w \rangle^2 \Bigr] \Bigl[
   \langle {\cal W}_{\rm p}\rangle + \langle {\cal W}_{\rm s}\rangle \Bigr]^{-1}.
\label{A4}
\end{eqnarray}
Also, the formula (\ref{15}) for the noise-reduction-parameter $ R_c $ of
photocounts can be appropriately modified:
\begin{eqnarray}  
 R_{c} &=& 1 + \Bigl[ (\eta_{\rm s} - \eta_{\rm i})^2 \bigl[
   \langle (\Delta {\cal W}_{\rm p})^2\rangle + K N(N-1) \langle {\cal W}_{\rm p}^w \rangle^2
   \bigr]  \nonumber \\
 & & - 2\eta_{\rm s}\eta_{\rm i} \langle {\cal W}_{\rm p}\rangle
 + \eta_{\rm s}^2 \langle (\Delta {\cal W}_{\rm s})^2\rangle
    + \eta_{\rm i}^2 \langle (\Delta {\cal W}_{\rm i})^2\rangle \Bigr]
     \nonumber \\
 & & \times \Bigl[ (\eta_{\rm s} + \eta_{\rm i}) \langle {\cal W}_{\rm p}\rangle
   + \eta_{\rm s} \langle {\cal W}_{\rm s}\rangle +
     \eta_{\rm i} \langle {\cal W}_{\rm i}\rangle \Bigr]^{-1}.
\label{A5}
\end{eqnarray}
We note that the corresponding quantities characterizing photon-number fields
are also given by formulas (\ref{A4}) and (\ref{A5}) in which we set $
\eta_{\rm s} = \eta_{\rm i} = 1 $.

The Fano factors $ F_{c,{\rm i}} $ and $ F_{n,{\rm i}} $ and the effective
detection efficiencies $ \eta^{\rm eff}_{\rm s} $ and $ \eta^{\rm eff}_{\rm i}
$, that show greater declinations between the experimental data and predictions
of the original model without the correlations in pump-pulse intensity
fluctuations for greater numbers $ N $ of grouped detection windows, are
plotted in Fig.~\ref{fig3}(b) and Fig.~\ref{fig5}(a), respectively. The solid
plain curves close to the experimental points in Figs.~\ref{fig3}(b) and
\ref{fig5}(a) drawn for the model with correlations in intra pump-pulse
intensity fluctuations and $ K \langle {\cal W}_{\rm p}^w \rangle^2 = 1\times
10^{-5} $ describe well the behavior of the studied quantities for great
numbers $ N $ of grouped detection windows. We have in the model $ \langle
{\cal W}_{\rm p}^w \rangle = m_{\rm p}b_{\rm p} = 0.10185 $ and so $ K = 0.965
\times 10^{-3} $, in accordance with the values of coefficients $ \bar{K}_{\rm
s} $ and $ \bar{K}_{\rm i} $ plotted in Fig.~\ref{fig5}(c). As it follows from
Eq.~(\ref{A5}), the correlations in pump-pulse intensity fluctuations influence
only weakly the values of noise-reduction-parameters $ R_c $ and $ R_n $ in our
case in which $ \eta_{\rm s} \approx \eta_{\rm i} $. We note that, in our
opinion, the slightly larger theoretical values of effective detection
efficiencies $ \eta^{\rm eff}_{\rm s} $ and $ \eta^{\rm eff}_{\rm i} $ in
Fig.~\ref{fig5}(a) in the area around $ N=100 $ indicate longer-range
death-time effects in the response of the used APDs and processing electronics
\cite{Straka2020}.


%

\end{document}